\documentstyle[12pt,titlepage]{article}
\begin{document}
\pagestyle{empty}
\baselineskip=0.212in

\begin{flushleft}
\large
{SAGA-HE-67-94  \hfill August, 1994}  \\
\end{flushleft}

\vspace{1.2cm}

\begin{center}

\Large{{\bf SU(2)-Flavor-Symmetry Breaking }} \\

\vspace{0.3cm}

\Large{{\bf in Nuclear Antiquark Distributions }} \\

\vspace{1.0cm}

\Large
{S. Kumano $^*$ }         \\

\vspace{0.6cm}

\Large
{Department of Physics, Saga University, Saga 840, Japan } \\

\vspace{0.1cm}

\Large{and}    \\

\vspace{0.1cm}

\Large
{The ECT$^\star$, Villa Tambosi, I-38050 Villazzano, Trento, Italy}         \\

\vspace{1.0cm}

\Large{ABSTRACT}

\end{center}

SU(2)-flavor-symmetry breaking in antiquark distributions
of the nucleon was suggested by the New Muon Collaboration (NMC)
in deep inelastic muon scattering.
As an independent test, Drell-Yan data for the tungsten target
have been used for examining the asymmetry.
We investigate whether there exists significant modification
of the $\bar u -\bar d$ distribution in nuclei
in a parton recombination model.
It should be noted that a finite $\bar u-\bar d$ distribution
is theoretically possible in nuclei
even if the sea is $SU(2)_f$ symmetric in the nucleon.
In neutron-excess nuclei such as the tungsten,
there exist more $d$-valence quarks than $u$-valence quarks, so that
more $\bar d$-quarks are lost than $\bar u$-quarks are due
to parton recombinations in the small $x$ region.
Our results suggest that the nuclear modification
in the tungsten is
a 2--10 \% effect on the $\bar u-\bar d$ distribution suggested by
the NMC data.
Nuclear effects on the flavor asymmetric distribution
could be an interesting topic for future theoretical and
experimental investigations.

\vspace{1.0cm}

\vfill

\noindent
{\rule{6.cm}{0.1mm}} \\

\vspace{-0.4cm}

\noindent
\normalsize
{$*$ ~Email: kumanos@himiko.cc.saga-u.ac.jp.} \\

\vspace{-0.2cm}
\hfill
{submitted for publication}

\vfill\eject
\pagestyle{plain}

\noindent
{\Large\bf {1. Introduction}}
\vspace{0.4cm}

It became possible to test the Gottfried sum rule \cite{GOTT} recently
because accurate experimental data could be taken in the small $x$ region.
This sum rule is a phenomenological one based on a naive parton model.
This is because we need a special assumption of the SU(2)-flavor-symmetric
antiquark distributions.
Therefore, even if the sum rule is broken, the QCD itself is not in danger.
However, it implies an interesting physics mechanism which we cannot
expect in the naive parton model.

The sum rule is described in the parton model as follows.
Integrating the proton and neutron structure-function difference
over $x$ with the isospin-symmetry assumption
($u_p=d_n\equiv u$, $d_p=u_n\equiv d$,
 $\bar u_p=\bar d_n\equiv \bar u$,
 $\bar d_p=\bar u_n\equiv \bar d$),
we obtain
$$
S_G ~\equiv~ \int_0^1 {{dx} \over x} [F_2^{\mu p}(x)-F_2^{\mu n}(x)]
{}~~.
\eqno{(1)}
$$
If the antiquark distributions are SU(2)$_{flavor}$ symmetric,
the second term vanishes and we obtain the Gottfried sum rule
($S_G=1/3$).

Although there is some indication of the sum-rule breaking
in the old SLAC data, it is rather surprising that
the New Muon Collaboration (NMC) found large violation.
They obtained the integral at $Q^2$=4 GeV$^2$ as \cite{NMC}
$$
\int_{0.004}^{0.8}{{dx} \over x} [F_2^{\mu p}(x)-F_2^{\mu n}(x)]=
0.221 \pm 0.008 \pm 0.019
{}~~.
\eqno{(2)}
$$
Adding contributions from the unmeasured region, they found
a significant deviation from the Gottfried sum rule,
$$
S_G ~=~ 0.235 \pm 0.026
{}~~.
\eqno{(3)}
$$
This NMC result indicates
that antiquark distributions are not SU(2)$_f$ symmetric
and we have $\bar d$ excess over $\bar u$.

This conclusion is contrary to the naive quark model expectation,
the SU(2)$_f$ symmetric sea ($\bar u=\bar d$).
The symmetric distributions are expected
because the sea is thought to be created
perturbatively through a gluon splitting into
a light quark pair, $u\bar u$ or $d\bar d$.
Thus, the NMC result suggests a nonperturbative mechanism for
explaining the symmetry breaking.
There are theoretical candidates such as
Pauli blocking models \cite{PAULI}, mesonic models \cite{MESON},
and others \cite{OTHERS}.

We do not discuss these theoretical models in this paper.
In spite of the NMC's claim of the flavor-symmetry breaking,
there is still possibility that the result could be explained
with the symmetric sea $\bar u=\bar d$.
This is because the smallest $x$ in the NMC experiment is 0.004
and the Gottfried sum rule may receive a significant contribution
from the very small $x$ region.
In order to test this problem, we should wait at least several
years for a possible HERA experiment at small $x$.
Therefore, it is desirable that we have independent experimental
evidence for the asymmetric antiquark distributions \cite{PIET}.

A possible way is to use Drell-Yan processes \cite{DYW}.
In fact,  the Drell-Yan data for the tungsten target have
been analyzed. At this stage, we cannot draw
a strong conclusion from the tungsten data whether
the light-quark-sea is symmetric or not.
On the other hand, nuclear effects are possibly significant
because the tungsten is a heavy nucleus.
If the nuclear modification is very large, the Drell-Yan analysis
cannot be directly compared with the NMC result.

The purpose of this paper is to investigate
a possible mechanism of producing a $\bar u-\bar d$
distribution in nuclei, especially in the tungsten nucleus.
In the near future, the Drell-Yan data for the proton and the deuteron
targets will be taken at Fermilab, so that our prediction can
be tested.
There is also a recent Drell-Yan result by the CERN-NA51 collaboration,
and it indicates a strong asymmetry
$\bar u/\bar d= 0.51 \pm 0.04 \pm 0.05$ at $x=0.18$ and $Q^2=20$ GeV$^2$
\cite{CERNDY}.
If accurate experimental results are supplied by
these Drell-Yan experiments for the nucleon and also for nuclear targets,
nuclear modification of the $\bar u -\bar d$ distribution
could become an interesting topic.
This paper is intended to shed light on such modification.

\vfill\eject

\noindent
{\Large\bf {2. Nuclear effects on the $\bar u-\bar d$ distribution}}
\vspace{0.4cm}

We investigate the SU(2)-flavor-asymmetric distribution $(\bar u-\bar d)_A$
in nuclei, especially in the tungsten nucleus.
We first show ``conventional" expectations without nuclear effects.
If there is no asymmetry in the nucleon ($\bar u=\bar d$) and
no nuclear modification, the distribution is symmetric
[$(\bar u-\bar d)_A=0$] as shown by the solid line in Fig. 1.
However, if there is asymmetry in the nucleon, the distribution
becomes
$$
x[\bar u(x)-\bar d(x)]_A = - \varepsilon x [\bar u(x)-\bar d(x)]_{proton}
{}~~,
\eqno{(1)}
$$
without considering nuclear modification.
The neutron-excess parameter $\varepsilon$ is defined by
$$
\varepsilon ={{N-Z} \over {N+Z}}  ~~~.
\eqno{(2)}
$$
This parameter is defined so that it satisfies $\varepsilon=0$
for isoscalar nuclei ($N=Z$) and it does $\varepsilon=1$ for neutron matter.
For example, it is 0.196 for the tungsten $_{74}^{184} W_{110}$.
Isospin symmetry is assumed in the parton distributions
of the proton and the neutron in Eq. (1).
Using the MRS-D0 parton distributions \cite{MRS93}, which were
obtained so as to reproduce the NMC data, we get the $\bar u-\bar d$
distribution in the tungsten nucleus as shown by the dashed curve
in Fig. 1.

Next, we address ourselves to nuclear modification
of the distributions in Fig. 1.
We discuss a mechanism of producing the asymmetric
distribution $(\bar u-\bar d)_{_A}$ in nuclei
even though antiquark distributions
are flavor symmetric in the nucleon [$(\bar u-\bar d)_N=0$].
Because antiquark distributions are dominant in the small
$x$ region, we should first find a mechanism of
nuclear shadowing for investigating nuclear antiquark distributions.
There are a number of theoretical ideas for explaining
the shadowing. We employ the parton-recombination model
in Ref. \cite{SKSHADOW} simply because the model can explain
many existing data of $F_2^A(x,Q^2)$.

In the parton picture, partons in different nucleons
could interact in a nucleus.
These interactions are especially important at small $x$
with the following reason.
In an infinite momentum frame, the average longitudinal
nucleon separation in a Lorentz contracted nucleus is
L=(2 fm)$M_A/P_A$=(2 fm)$m_{_N}/p_{_N}$, and the longitudinal
localization size of a parton with momentum $xp_{_N}$ is
$\Delta L=1/(xp_{_N})$. If the parton dimension exceeds
the average separation ($\Delta L > L$) in the small $x$ region ($x<0.1$),
partons from different nucleons could interact.
This is an extra effect which does not exist in a single nucleon.
The interaction is called parton recombination or parton fusion.
In discussing antiquark distributions in nuclei, this effect should
be taken into account properly.
The recombination contributions to an antiquark distribution are given
in the appendix.
There are three physics ingredients in Eq. (A.3).
The first and the second integrals are from interactions of
an antiquark $\bar q_i$ in the nucleon $n1$ with
a gluon in the nucleon $n2$;
the third and the sixth integrals are from those of
an antiquark $\bar q_i$ in the nucleon $n1$ with
a quark $q_i$ in the nucleon $n2$;
the forth and the fifth integrals are from those of
a gluon in the nucleon $n1$ with
an antiquark $\bar q_i$ in the nucleon $n2$.

The lengthy equation becomes simpler
if we assume that antiquark distributions in the nucleon
are $SU(2)_f$ symmetric ($\bar u=\bar d$)
and that the leak-out quark ($q^*$) is a sea quark with
the $SU(2)_f$ symmetry ($u_{sea}^*=d_{sea}^*$).
Obviously, most terms in $[\bar u(x)-\bar d(x)]_A$
vanish except for the last integral.
Under these assumptions,
modification of the flavor asymmetry in a nucleus becomes
$$
x[\Delta \bar u(x) - \Delta \bar d(x)]_A=
\varepsilon
{ {4K} \over 9 } x \int_0^1 dx_2   ~ x \bar u^* (x)
        ~ x_2 [u_v(x_2)-d_v(x_2)]~  {{x^2+x_2^2} \over {(x+x_2)^4}}
{}~~,
\eqno{(3)}
$$
where $u_v(x)$ and $d_v(x)$ are u and d
valence-quark distributions in the proton.
Now the meaning of creating the flavor asymmetry in nuclei becomes
clear.
In a neutron-excess nucleus ($\varepsilon >0$),
the $d_v$ quark number is larger than the $u_v$ quark one.
Hence, more $\bar d$ quarks are lost than $\bar u$ quarks
in the parton recombination process
$\bar q q \longrightarrow G$.
This situation is illustrated in Fig. 2.
The modification of the flavor-asymmetric distribution
is directly proportional to
the neutron-excess parameter.
Hence, the nuclear modification is $SU(2)_f$ symmetric
in isoscalar nuclei ($\varepsilon=0$), and it becomes larger
as the neutron excess increases.

We evaluate Eqs. (3) and (A.3) with the input
parton distributions MRS-D0 (1993) \cite{MRS93},
$Q^2$=4 GeV$^2$, $\Lambda$=0.2 GeV in $\alpha_s$,
$n_f$=3 (number of flavor), and $z_0$=2 fm
(cutoff for parton leaking, see Ref. \cite{SKSHADOW} for details)
in the tungsten nucleus $_{74}^{184} W_{110}$ ($\varepsilon=0.196$).
In evaluating Eq. (3), the $SU(2)_f$ symmetric sea in the nucleon
is used by setting $\Delta=0$ in the MRS-D0 distribution.
Obtained results are shown in Fig. 3, where the solid curve shows
the $x[\Delta\bar u-\Delta\bar d]_A$ distribution (per nucleon)
of the tungsten nucleus
in Eq. (3) with the $SU(2)_f$ symmetric sea in the nucleon,
and the dashed curve shows the one in Eq. (A.3)
with the $SU(2)_f$ asymmetric sea distributions as the input.

As expected in a neutron-excess nucleus, the parton recombinations
produce a finite $SU(2)_f$-breaking antiquark distribution even if
antiquark distributions are $SU(2)_f$ symmetric in the nucleon.
Furthermore, it is a positive contribution to $\bar u-\bar d$
because of the d-valence-quark excess over u-valence
in the neutron-excess nucleus.
We briefly comment on $Q^2$ dependence of our calculation.
Even though explicit $Q^2$ dependence is not shown in Eq. (3),
the dependence is included in the factor $K$ and
in parton distributions $p(x,Q^2)$.
Because the factor $K$ is proportional to $\alpha_s(Q^2)/Q^2$,
the nuclear flavor asymmetry may seem to be very large at small $Q^2$.
However, the quark distribution $u_v(x)-d_v(x)$ in Eq. (3)
becomes very small in the small $x$ region, so that the overall
$Q^2$ dependence is not so significant.
There are merely factor-of-two differences
between the asymmetric distribution
at $Q^2$=4 GeV$^2$ and the one at $Q^2 \approx 1$ GeV$^2$.

The situation is changed if the $SU(2)_f$ asymmetric distribution
is used as the input distribution. In this case, all the process
in Eq. (A.3) contribute and others become as large as
the $q \bar q \rightarrow G$ contribution.
Because pn (proton-neutron) and np recombination contributions cancel,
the flavor asymmetry is given by
$$
x[\Delta \bar u(x) - \Delta \bar d(x)]_A =
-(w_{nn} - w_{pp})
x[\Delta \bar u(x) - \Delta \bar d(x)]_{pp}
{}~~,
\eqno{(4)}
$$
where $w_{nn}$ and $w_{pp}$ are neutron-neutron and proton-proton
recombination probabilities in Eq. (A.2).
Their difference is equal to the neutron-excess parameter,
$w_{nn} - w_{pp}=\varepsilon$.
$[\bar u(x)-\bar d(x)]_{pp}$ is the asymmetry produced in the
proton-proton recombination.
The $q\bar q \rightarrow G$ contribution in the sixth integral
of Eq. (A.3) to $[\bar u(x)-\bar d(x)]_A$ is positive as we found
in the $SU(2)_f$-symmetric input case.
However, it is partly canceled by
the $q\bar q \rightarrow G$ contribution in the third integral
due to the input-sea-quark asymmetry.
Furthermore, $\bar q G \rightarrow \bar q$ contributions
in the second and the fifth integrals are larger than the sixth integral
at small $x$,
and they are negative due to the $\bar d$-excess over $\bar u$ in the proton.
Summing up these contributions, we find $\bar d$-excess over $\bar u$
in the small $x$ region ($x<0.01$)
as shown in Fig. 3.
On the contrary, we find $\bar u$-excess over $\bar d$ in the larger $x$
region ($x>0.05$) because $\bar q G \rightarrow \bar q$ processes
in the first and the fourth integrals dominate the contribution.
Without considering these nuclear effects,
we obtained
the asymmetry in the tungsten nucleus in Fig. 1
($Max[x\bar u(x)-x\bar d(x)]_A\approx +0.005$).
The nuclear modifications are shown in Fig. 3
($Max[x\bar u(x)-x\bar d(x)]_A\approx +0.0001 ~or~ +0.00025$).
Considering the factor of two coming from the $Q^2$ dependence,
we find that the nuclear modification is of the order of 2\%--10\%
compared with the asymmetry suggested by the MRS-D0 distribution.
In this way, even though nuclear modification is rather small,
the $SU(2)_f$ asymmetry in antiquark distributions
is very interesting quantity for testing underlying nuclear dynamics.

\vfill\eject
\noindent
{\Large\bf {3. Conclusions}}
\vspace{0.4cm}

We investigated nuclear effects on the $SU(2)$-flavor-symmetry
in antiquark distributions.
As a result, it is interesting to find that
a finite flavor-breaking distribution
in a nucleus ($[\bar u(x)-\bar d(x)]_A \ne 0$) is possible
even though it is symmetric
in the nucleon ($\bar u(x)-\bar d(x)=0$).
According to our recombination model, the nuclear effects
on $[\bar u(x)-\bar d(x)]_A$
are of the order of 2\%--10\% compared with
the one estimated by the NMC flavor asymmetry in the nucleon.
Because the Drell-Yan experiments on the proton and deuteron targets
will be done at Fermilab in the near future,
it is in principle possible
to study the nuclear modification experimentally.
The nuclear effects are important for testing underlying
nuclear dynamics, and studies of these nuclear effects may
help to understand the physics origin of the flavor asymmetry
in the nucleon.
The author hopes that this paper motivates theorists and experimentalists
for investigating further details of nuclear $\bar u-\bar d$ distributions.

$~~~$

$~~~$

\noindent
{\Large\bf {Acknowledgment}}
\vspace{0.4cm}

This research was partly supported by the Grant-in-Aid for
Scientific Research from the Japanese Ministry of Education,
Science, and Culture under the contract number 06640406.
S.K. thanks the European Centre for Theoretical Studies
in Nuclear Physics and Related Areas (ECT$^\star$) in Trento
for its hospitality and for partial support for this project.

\vfill\eject
\noindent
{\Large\bf {Appendix}}
\vspace{0.4cm}

Parton recombination effects on the antiquark distribution
$\bar q_i(x)$ are given by
$$
x \cdot \Delta \bar q_{i,A} (x) ~=~ w_{pp} x \cdot \Delta \bar q_{i,pp} (x)
                                ~+~ w_{pn} x \cdot \Delta \bar q_{i,pn} (x)
                                ~+~ w_{np} x \cdot \Delta \bar q_{i,np} (x)
                                ~+~ w_{nn} x \cdot \Delta \bar q_{i,nn} (x)
{}~~,
\eqno{(A.1)}
$$
where
$$
 w_{pp}= {{Z(Z-1)} \over {A(A-1)}}~, ~~
               w_{pn}= {{ZN} \over {A(A-1)}}~, ~~
               w_{np}= {{NZ} \over {A(A-1)}}~, ~~
               w_{nn}= {{N(N-1)} \over {A(A-1)}}
{}~~,
\eqno{(A.2)}
$$
are the combination probabilities of two nucleons.
For example, $w_{pp}$ is the probability of a proton-proton combination.
$\Delta \bar q_{i,n1 \hspace{0.05cm} n2} (x)$ is the modification
of the antiquark distribution with flavor $i$ due to a parton interaction
in the nucleon $n1$ with a parton in the nucleon $n2$.
The details of the recombination mechanism is in Ref. \cite{SKSHADOW}.
We simply employ the recombination result
$$
x \cdot \Delta \bar q_{i,n1 \hspace{0.05cm} n2} (x) ~=~
  +   { K \over 6 } \int_0^x {{dx_2} \over {x_2}}
     ~   \biggl[~ (x-x_2) \bar q_{i,n1} (x-x_2)
     ~   \bigl\{1+( {{x-x_2} \over x} )^2 \bigr\}
$$

\vspace{0.3cm}
\noindent
$
{}~~~~~~~~~~~~~~
{}~~~~~~~~~~~~~~
{}~~~~~~~~~~~~~~~
\displaystyle{  - ~x \bar q_{i,n1} (x)~ {x \over {x+x_2}}~
                       \bigl\{ 1+({x \over {x+x_2}} )^2 \bigr\}~ \biggr]
                      ~x_2 G^*_{n2} (x_2)~
}
$

\vspace{0.3cm}
\noindent
$
{}~~~~~~~~~~~~~~~~~~~~~~~~
\displaystyle{
- { K \over 6 } \int_x^1 {{dx_2} \over {x_2}} ~x \bar q_{i,n1} (x)~
         x_2 G^*_{n2} (x_2) ~{ x \over {x+x_2}}~
         \bigl\{ 1+( {x \over {x+x_2}} )^2 \bigr\}
}
$

\vspace{0.3cm}
\noindent
$
{}~~~~~~~~~~~~~~~~~~~~~~~~
\displaystyle{
- { {4K} \over 9 } x\int_0^1 dx_2  ~x \bar q_{i,n1}(x)~
         x_2  q_{i,n2}^*(x_2)~ {{x^2+x_2^2} \over {(x+x_2)^4}}
}
$

\vspace{0.3cm}
\noindent
$
{}~~~~~~~~~~~~~~~~~~~~~~~~
\displaystyle{
+  { K \over 6 } \int_0^x {{dx_1} \over {x_1}}
       ~  x_1 G_{n1} (x_1)~ \biggl [~ (x-x_1) \bar q_{i,n2}^*(x-x_1) ~
        \bigl \{ 1+( {{x-x_1} \over x} )^2 \bigr \}
}
$

\vspace{0.3cm}
\noindent
$
{}~~~~~~~~~~~~~~
{}~~~~~~~~~~~~~~
{}~~~~~~~~~~~~~~
{}~~~~~~~~~~~~~~~~~~
\displaystyle{
              -~x \bar q_{i,n2}^* (x)~ {x \over {x+x_1}}~
                   \bigl\{  1+({x \over {x+x_1}})^2 \bigr\} ~ \biggr ]
}
$

\vspace{0.3cm}
\noindent
$
{}~~~~~~~~~~~~~~~~~~~~~~~~
\displaystyle{
- { K \over 6 } \int_x^1 {{dx_1} \over {x_1}}
        ~ x_1 G_{n1} (x_1) ~x \bar q_{i,n2}^*(x)~ { x \over {x+x_1}}~
         \bigl \{ 1+( {x \over {x+x_1}} )^2 \bigr \}
}
$

\vspace{0.3cm}
\noindent
$
{}~~~~~~~~~~~~~~~~~~~~~~~~
\displaystyle{
- { {4K} \over 9 } x \int_0^1 dx_2   ~ x \bar q_{i,n1}^* (x)
        ~ x_2 q_{i,n2}(x_2)~  {{x^2+x_2^2} \over {(x+x_2)^4}}
{}~~~~
}
{}~~.
\hfill (A.3)
$

\vspace{0.5cm}
\noindent
The asterisk mark indicates a leak-out parton and
$K$ is given as $K=9A^{1/3}\alpha_s(Q^2)/(2R_0^2Q^2)$
with $R_0=1.1$ fm.

\vfill\eject

\vspace{3.0cm}
\noindent
{\Large\bf{Figure Captions}} \\

\vspace{-0.38cm}
\begin{description}
   \item[Fig. 1]
The dashed curve shows the
$x[\bar u(x)-\bar d(x)]_A$ distribution in the tungsten nucleus
according to the MRS-D0 distribution.
No nuclear modification is considered. Because of the neutron excess,
$\bar u(x)$ is larger than
$\bar d(x)$ on the contrary to the proton case.
If there is no flavor asymmetry in the nucleon and
no nuclear modification, we obtain no flavor asymmetry in the nucleus
as shown by the solid line.
   \item[Fig. 2]
Schematic pictures of parton-recombination processes.
Due to d-valence-quark excess over u-valence
in a neutron-excess nucleus,
more $\bar d$ quarks are lost than $\bar u$ quarks are
in the recombination process.
   \item[Fig. 3]
Parton-recombination effects on the flavor distribution
$x[\bar u(x)-\bar d(x)]_A$ in the tungsten nucleus.
The solid curve is calculated by using $SU(2)_f$ symmetric
antiquark distributions and the dashed one is by $SU(2)_f$ asymmetric
antiquark distributions given by the MRS-D0 parametrization.
It is interesting to find a finite $\bar u-\bar d$
distribution even if antiquark distributions
are $SU(2)_f$ symmetric in the nucleon.

\end{description}

\end{document}